\documentclass[conference,compsoc]{IEEEtran}
\ifCLASSOPTIONcompsoc
  \usepackage[nocompress]{cite}
\else
  \usepackage{cite}
\fi
\ifCLASSINFOpdf
  \usepackage[pdftex]{graphicx}
\else
  \usepackage[dvips]{graphicx}
\fi
\usepackage{amsmath}
\usepackage{array}

\ifCLASSOPTIONcompsoc
  \usepackage[caption=false,font=footnotesize,labelfont=sf,textfont=sf]{subfig}
\else
  \usepackage[caption=false,font=footnotesize]{subfig}
\fi
\usepackage{url}

\usepackage{xcolor}
\usepackage{verbatim}
\usepackage{hyperref}
\usepackage{listings}
\usepackage{multicol}
\usepackage{tikz}
\usetikzlibrary{automata,positioning,arrows, matrix}
\usepackage{standalone}
\usepackage{floatrow}
\usepackage{import}

\usepackage{listings, xcolor}

\definecolor{verylightgray}{rgb}{.97,.97,.97}

\lstdefinelanguage{Solidity}{
	keywords=[1]{anonymous, assembly, assert, balance, break, call, callcode, case, catch, class, constant, continue, constructor, contract, debugger, default, delegatecall, delete, do, else, emit, event, experimental, export, external, false, finally, for, function, gas, if, implements, import, in, indexed, instanceof, interface, internal, is, length, library, log0, log1, log2, log3, log4, memory, modifier, new, payable, pragma, private, protected, public, pure, push, require, return, returns, revert, selfdestruct, send, solidity, storage, struct, suicide, super, switch, then, this, throw, transfer, true, try, typeof, using, value, view, while, with, addmod, ecrecover, keccak256, mulmod, ripemd160, sha256, sha3}, 
	keywordstyle=[1]\color{blue}\bfseries,
	keywords=[2]{address, bool, byte, bytes, bytes1, bytes2, bytes3, bytes4, bytes5, bytes6, bytes7, bytes8, bytes9, bytes10, bytes11, bytes12, bytes13, bytes14, bytes15, bytes16, bytes17, bytes18, bytes19, bytes20, bytes21, bytes22, bytes23, bytes24, bytes25, bytes26, bytes27, bytes28, bytes29, bytes30, bytes31, bytes32, enum, int, int8, int16, int24, int32, int40, int48, int56, int64, int72, int80, int88, int96, int104, int112, int120, int128, int136, int144, int152, int160, int168, int176, int184, int192, int200, int208, int216, int224, int232, int240, int248, int256, mapping, string, uint, uint8, uint16, uint24, uint32, uint40, uint48, uint56, uint64, uint72, uint80, uint88, uint96, uint104, uint112, uint120, uint128, uint136, uint144, uint152, uint160, uint168, uint176, uint184, uint192, uint200, uint208, uint216, uint224, uint232, uint240, uint248, uint256, var, void, ether, finney, szabo, wei, days, hours, minutes, seconds, weeks, years},	
	keywordstyle=[2]\color{teal}\bfseries,
	keywords=[3]{block, blockhash, coinbase, difficulty, gaslimit, number, timestamp, msg, data, gas, sender, sig, value, now, tx, gasprice, origin},	
	keywordstyle=[3]\color{violet}\bfseries,
	identifierstyle=\color{black},
	sensitive=false,
	comment=[l]{//},
	morecomment=[s]{/*}{*/},
	commentstyle=\color{gray}\ttfamily,
	stringstyle=\color{red}\ttfamily,
	morestring=[b]',
	morestring=[b]"
}

\definecolor{darkblue}{rgb}{0,0,0.5}
\definecolor{darkgreen}{rgb}{0,0.5,0}
\newcommand{\var}[1]{\mathtt{#1}}
\newcommand{\true}{\mathsf{True}}

\begin{document}
%
\title{Smart Contract Design Meets State Machine Synthesis: Case Studies}

\author{\IEEEauthorblockN{Dmitrii Suvorov}
\IEEEauthorblockA{ITMO University\\
JetBrains Research\\
Email: dmsuvorov@corp.ifmo.ru}
\and
\IEEEauthorblockN{Vladimir Ulyantsev}
\IEEEauthorblockA{ITMO University\\
Email: ulyantsev@corp.ifmo.ru}}


%


\maketitle

\begin{abstract}
Modern blockchain systems support creation of smart contracts -- stateful programs hosted and executed on a blockchain. Smart contracts hold and transfer significant amounts of digital currency which makes them an attractive target for security attacks. It has been shown that many contracts deployed to public ledgers contain security vulnerabilities. Moreover, the design of blockchain systems does not allow the code of the smart contract to be changed after it has been deployed to the system. Therefore, it is important to guarantee the correctness of smart contracts prior to their deployment.

Formal verification is widely used to check smart contracts for correctness with respect to given specification. In this work we consider program synthesis techniques in which the specification is used to generate correct-by-construction programs. We focus on one of the special cases of program synthesis where programs are modeled with finite state machines (FSMs). We show how FSM synthesis can be applied to the problem of automatic smart contract generation. Several case studies of smart contracts are outlined: crowdfunding platform, blinded auction and a license contract. For each case study we specify the corresponding smart contract with a set of formulas in linear temporal logic (LTL) and use this specification together with test scenarios to synthesize a FSM model for that contract. These models are later used to generate executable Solidity code which can be directly used in a blockchain system.

\end{abstract}


%
\IEEEpeerreviewmaketitle

\section{Introduction}
Since the invention of a blockchain data structure in 2008 various
cryptocurrencies have been emerging, evolving and gaining popularity.
This popularity is explained by the fact that blockchain systems are
fully operable without a trusted entity. Recent cryptocurrencies
support creation of \textit{smart contracts} -- stateful programs
executed on a blockchain that encode the rules governing
transactions. The execution of smart contracts is enforced by the
consensus algorithm in the underlying blockchain system.

Smart contracts are a powerful tool to encode arbitrary contractual
agreements in a machine-readable form but as with any programs they
are error-prone and hard to reason about. Furthermore, the blockchain
systems design principles make impossible contract's code modification
after it has been deployed to blockchain.  Also smart contracts hold
and transfer significant amount of digital currency which makes them
an attractive target of various attacks and drastically increases the
cost of an error in smart contract code. It has been shown that many
contracts deployed to public ledgers contain security vulnerabilities
and these vulnerabilities have led to theft of millions of US dollars
in cryptocurrency equivalent. Thus, it is of paramount importance to
ensure that smart contracts are correct. Various methods based on
formal verification have been proposed to achieve this
goal~\cite{bhargavan2016formal, luu2016making, hirai2017defining}.

Another method to build correct programs is program synthesis, which
has a lot in common with formal verification.  The problem of program
synthesis is formulated as follows: given a specification in formal
logic, construct a program conforming to that specification. This
problem is known to be undecidable in general, however various methods
were proposed for some special cases of programs. Unlike formal
verification, automated synthesis of smart contracts has received very
little attention, although program synthesis application for smart
contracts looks promising given the fact that they are relatively
small (less than 100 SLOC in average~\cite{hegedus2019towards}).

Program synthesis is a very broad topic and in this work we only focus
on the problem of FSM synthesis where programs are defined in terms of
FSMs. The rationale behind this is that smart contract logic can often
be expressed with an FSM. Moreover, modeling contracts as FSMs is a
recommended design pattern for Solidity -- a language of
Ethereum~\cite{wood2014ethereum}
contracts~\cite{solidity-fsm-pattern}. In fact there is a tool
\textsc{VeriSolid}~\cite{mavridou2017designing} that facilitates
creating FSM smart contract models and generating Solidity code from
these models. A user can specify temporal properties and verify that
generated contracts conform to these properties.

\begin{figure}[h!]
  \centering
  \includegraphics[scale=0.6]{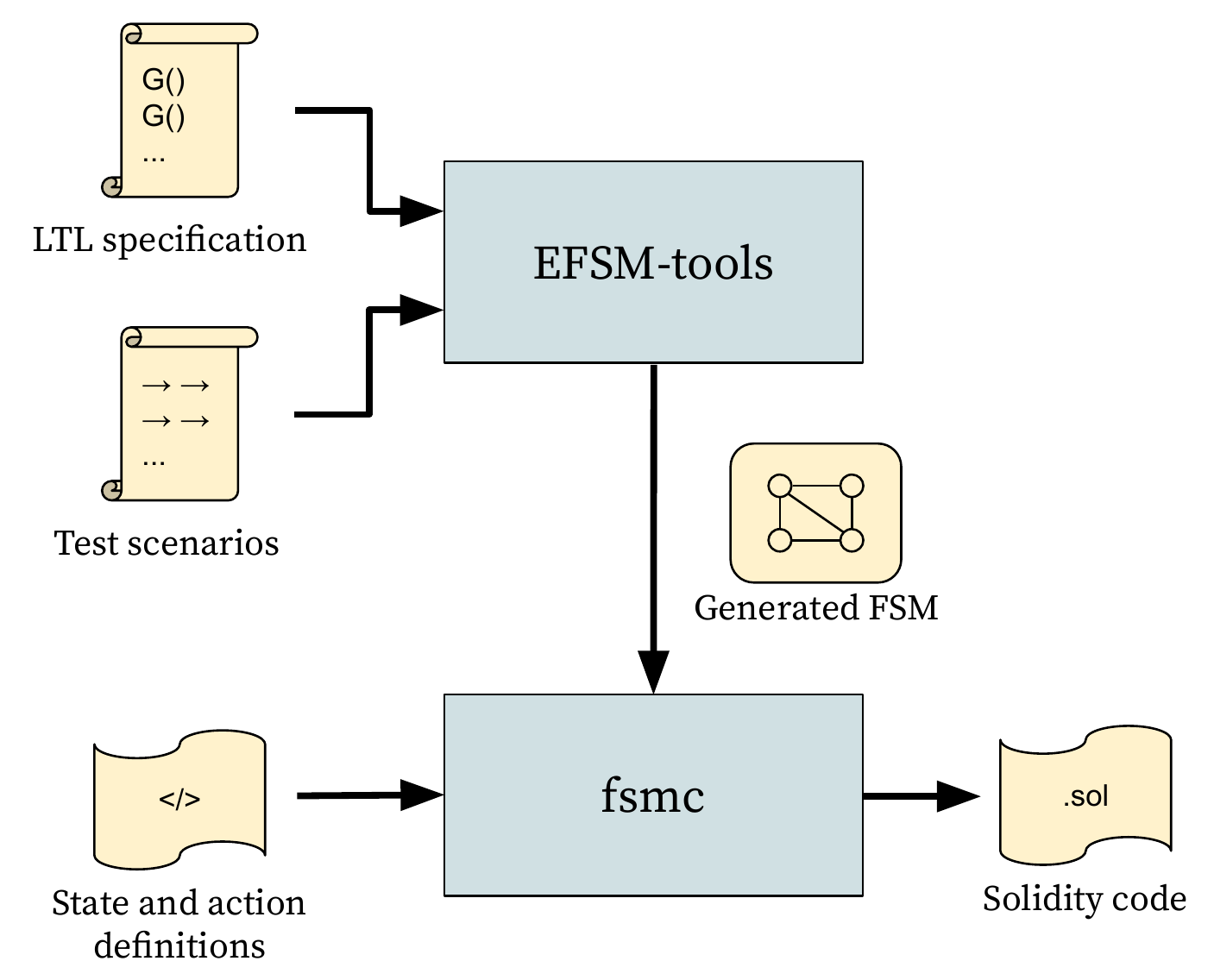}
  \caption{Data-flow diagram of the proposed approach.}
  \label{fig:approach-diagram}
\end{figure}

In this work we employ techniques and tools
(\textsc{EFSM-tools}~\footnote{\url{https://github.com/ulyantsev/EFSM-tools/}})
outlined in~\cite{efsm-tools}. We have no intent to compare different
FSM synthesis tools as for case studies in
Section~\ref{sec:case-studies} synthesis solving terminates within
seconds. In our approach input events of a FSM correspond to methods
of a contract and output actions -- to implementations of those
methods. We synthesize smart contract FSM models based on formal
specification in temporal logic and test scenarios.  Afterwards, given
the contract's state declaration and the implementation of its methods
in a corresponding programming language, we can generate contract's
code that is guaranteed to be correct with respect to the
specification.  Code generation is straightforward and similar to that
in \textsc{VeriSolid} tool, however \textsc{VeriSolid} does not
support multiple transitions labeled by the same event so we had to
implement our own tool
\textsc{fsmc}~\footnote{\url{https://github.com/d-suvorov/fsmc/}}.
The high-level data-flow diagram for the proposed approach is shown in
Figure~\ref{fig:approach-diagram}.

\textbf{Contributions.} To the best of our knowledge, this work is the
first attempt to employ specification-based program synthesis
techniques to automatically generate smart contract source
code. Specifically, we provide case studies to show that LTL synthesis
can be successfully applied for some types of contracts to generate
their FSM models and use these models to obtain source code that meets
some formal properties. We also briefly discuss the applicability of
program synthesis for automated smart contract generation.


 

\section{Background and related work}
In this section we introduce some basic concepts of blockchain systems
and smart contracts necessary to understand the rest of the paper.
Then we define FSMs that are used to model smart contracts and
introduce formalism that are used to specify them. Finally we state
the problem of specification-based FSM synthesis and provide some
references to its efficient solutions.


\subsection{Smart contracts}
A blockchain is a list of records, called blocks, containing some
data. Blockchain can be used as a ledger that is maintained in a
distributed network, for instance in cryptocurrencies this ledger
stores a transaction list. Effectively, a ledger in cryptocurrencies
stores the mapping from accounts to their balances in digital
currency. We refer to this digital currency as \textit{coins}.  The
nodes of a network called \textit{miners} execute consensus algorithm
and decide on which blocks will be added.  It is assumed that the
majority of nodes are honest as they are incentivized to add new valid
blocks, and the integrity of the system is based on that assumption.

Modern blockchain systems support \textit{smart contracts} --
executable programs stored on a blockchain.  A contract is executed by
miners which agree on the outcome of the execution and update the
blockchain accordingly. Hence arbitrary contractual agreements can be
expressed in program code and enforced without relying on a trusted
party. Most popular smart contract systems share the same concepts but
for the rest of the paper we consider Ethereum-like smart contracts.

In Ethereum~\cite{wood2014ethereum}, smart contracts are a type of
accounts associated with executable code and a storage file.  Smart
contracts can be created by sending a transaction of a special kind to
a blockchain.  A code of the contract consists of methods -- entry
points which are called when transactions are send to the address of
that contract. Essentially, transactions act as method
invocations. Contracts can receive coins with these transactions and
send coins to other accounts via \texttt{send} instructions. Each
instruction of a method consumes some amount of \textit{gas} during
execution. The user who sends a transaction must pay gas for its
execution. If a transaction runs out of gas during its execution the
control returns to sender. An example of a smart contract is shown in
the next section.

The problem of creating correct smart contracts have been actively
studied over the past years.  One of the first work by
Delmolino~\emph{et al.} outlines common pitfalls specific to smart
contract development. Since then, a variety of techniques have been
used to verify smart contracts.  Different tools based on symbolic
execution were created: \textsc{Oyente}~\cite{luu2016making},
\textsc{Mythril}~\cite{mythrill-github},
\textsc{Manticore}~\cite{manticore-github},
\textsc{Maian}~\cite{nikolic2018finding}.  An early work of
Bhargavan~\emph{et al.} uses F* programming
language~\cite{bhargavan2016formal}. Lately modern theorem provers
have been employed to formalize different aspects of smart contracts
in blockchain systems~\cite{hirai2017defining, amani2018towards,
  grishchenko2018semantic} and used to mechanize reasoning about those
aspects.  Sergey~\emph{et al.}~\cite{sergey2018scilla} design a new
functional language, implement its embedding into Coq and mechanize
proofs of safety and liveness properties of smart
contracts. Flint~\cite{schrans2018writing} is a programming language
that was designed specifically for writing robust smart
contracts. Flint employs linear type theory to prevent unintentional
loss of coins. An interesting example of contract-oriented languages
are Bamboo~\cite{bamboo-github} and
Obsidian~\cite{coblenz2017obsidian} as they model smart contracts as
state machines and make state transitions explicit.  Model checking
can also be used to verify smart contracts. Nehai~\emph{et
  al.}~\cite{nehai2018model} use \textsc{NuSMV} model checker to
create a blockchain application model (including a blockchain model
itself) and check its temporal properties.

Idelberger~\emph{et al.}~\cite{idelberger2016evaluation} propose to
use defeasible deontic logic to create smart contracts, which is
somewhat similar to our approach. However, the execution of such smart
contracts relies on an external logic engine. This setup negatively
affects the performance. We generate FSM models which can be encoded
in some programming language and executed directly.


\subsection{Specification-based FSM synthesis}

\begin{figure}[h!]
  

  \sidesubfloat[]{
  \begin{tikzpicture}
    [
      ->,>=stealth',
      shorten >=1pt,
      node distance=3.5cm,
      initial text=$ $
    ]

    \node[state,initial] (1) {$1$};
    \node[state,right of=1] (2) {$2$};

    \draw (1) edge[bend left, above] node{$e_2 / z_1$} (2)
          (2) edge[bend left, below] node[text width=1cm,align=center]{$e_1 / z_1$ \\ $e_2 / z_2$} (1)
          (1) edge[loop above] node{$e_1 / z_1$} (1);
  \end{tikzpicture}
  } \\
  \sidesubfloat[]{
  \begin{tikzpicture}
    [
      ->,>=stealth',
      shorten >=1pt,
      node distance=2.5cm,
      initial text=$ $,
      state/.style={
           rectangle,
           rounded corners,
           draw=black, thick,
           minimum height=2em,
           inner sep=2pt,
           text centered,
      },
      every loop/.style={min distance=10mm,looseness=10}
    ]

    \node[state,initial] (1) {$(1, e_1, z_1, 1)$};
    \node[state,right of=1,xshift=1cm] (2) {$(2, e_1, z_1, 1)$};
    \node[state,initial,below of=1] (3) {$(1, e_2, z_1, 2)$};
    \node[state,below of=2] (4) {$(2, e_2, z_2, 1)$};

    \draw
    (2) edge (1)
    (1) edge (3)
    (4) edge (1)
    (3) edge[bend left=8] (2)
    (2) edge[bend left=8] (3)
    (3) edge[bend left=7] (4)
    (4) edge[bend left=7] (3)
    (1) edge[loop above]  (1);
  \end{tikzpicture}
  }
  
  \caption{An example of an FSM (a) and its Kripke structure (b).}
  \label{fig:kripke}
\end{figure}
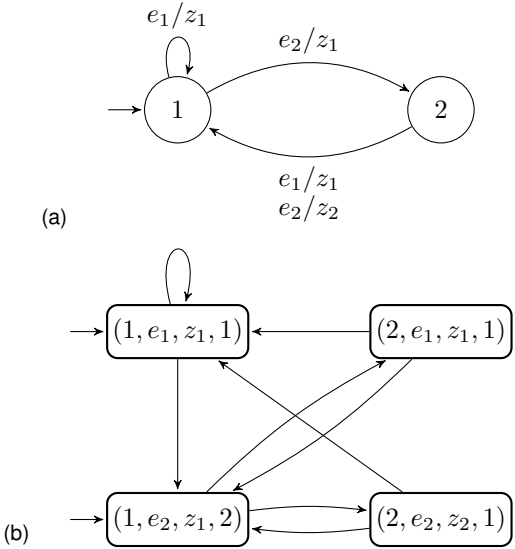

We are following~\cite{efsm-tools} and define a \textit{finite state
  machine} (FSM) as a tuple $(S, s_{\mathrm{init}}, E, Z, \delta,
\lambda)$, where
\begin{itemize}
  \item $S$ is a finite set of states,
  \item $s_{\mathrm{init}} \in S$ is the \textit{initial state},
  \item $E$ is a finite set of input \textit{events},
  \item $Z$ is a finite set of output \textit{actions},
  \item $\delta : S \times E \to S$ is the \textit{transition
    function},
  \item $\lambda : S \times E \to Z^{*}$ is the \textit{output
    function} (with $Z^{*}$ we denote a set of strings over $Z$).
\end{itemize}
An FSM reads a sequence of input events one by one and transforms it
into a sequence of output actions.  With each input event it generates
new output actions according to $\lambda$ and changes its active state
according to $\delta$.

\textbf{Model checking.}  Model checking is a technique for
automatically verifying finite-state systems with respect to a given
specification~\cite{grumberg1999model}. It is common to formalize the
specification as a formula in temporal logic. In linear temporal logic
(LTL), formulas express some properties of execution
paths~\cite{pnueli1977temporal}.  To proceed with its definition we
first define a Kripke structure. With $P$ we denote a set of atomic
propositions, which characterize execution states.  Formally, a Kripke
structure is a tuple $(S_K, I, T, L)$, where
\begin{itemize}
  \item $S_K$ is a set of states,
  \item $I \subseteq S_K$ is a set of initial states,
  \item $T \subseteq S_K \times S_K$ is a transition relation, which
    must be left-total (that is, from each state there is a transition
    to at least one state),
  \item $L : S_K \to 2^P$ is a labeling function.
\end{itemize}
An example of an FSM and a corresponding Kripke structure is shown in
Figure~\ref{fig:kripke}. To label transitions we use this notation:
\textit{input event / output action}.

LTL formulas are defined over infinite paths in Kripke structures. The
formulas are built up from temporal operators, atomic propositions and
connectives familiar from propositional logic ($\wedge$, $\vee$,
$\neg$, $\to$).  If $f$ is a Boolean formula, then it simply states
with which atomic propositions the first state of the path is marked.
If $f$ is an LTL formula, then saying that $f$ holds for a state of an
infinite path means that it holds for the infinite suffix of the path
starting from this state. The following temporal operators can be
used.
\begin{itemize}
  \item The $ne\textbf{X}t$ operator: $\textbf{X}\:f$ means that $f$
    has to hold at the next state of the path.
  \item The $\textbf{G}lobally$ operator: $\textbf{G}\:f$ means that
    $f$ has to hold on the entire suffix of the path.
  \item The $\textbf{F}uture$ operator: $\textbf{F}\:f$ means that $f$
    eventually has to hold (somewhere on the suffix of the path
    starting from this state).
  \item The $\textbf{U}ntil$ operator: $f\:\textbf{U}\:g$ means that
    $f$ has to hold at least until $g$ becomes true, which must hold
    at this or some future state.
  \item The $\textbf{R}elease$ operator: $f\:\textbf{R}\:g$ means that
    $g$ has hold until and including the point where $f$ first becomes
    true; if $f$ never becomes true, $g$ must remain true forever.
\end{itemize}
The formula $f$ is true for some Kripke model $M$ means that it is
satisfied for all infinite paths of $M$.



\textbf{FSM synthesis.} The problem of FSM synthesis by the
specification is well-known. In its different statements the
specification may be given as temporal formula, a set of test
scenarios or the combination of the two.  A test scenario for FSM $(S,
s_{init}, E, Z, \delta, \lambda)$ is a sequence of pairs $(e_1, A_1),
\dotsc, (e_n, A_n)$, where each $e_i \in E$ and $A_i \in Z^{*}$ and, a
FSM conforms to it if and only if it produces a sequence of actions
$A_1 \cdot A_2 \cdot \dotsc \cdot A_n$ (with $\cdot$ we denote
sequence concatenation) given a sequence of events $e_1, \dotsc, e_n$
as its input. Exact synthesis methods are mostly based on transition
to SAT~\cite{efsm-tools, walkinshaw}. In~\cite{efsm-tools} different
approaches based on transition to SAT and QSAT were examined. In the
most efficient approach scenarios are encoded in SAT and LTL formulas
are incorporated with iterative counterexample prohibition. In
\textsc{BoSy} tool~\cite{not-bosy, bosy}, encoding in QSAT instead of
SAT is used and a transition system is generated only from a set of
LTL formulas. The generated transition system is guaranteed to be
minimum in terms of the number of states.

\section{Case studies}\label{sec:case-studies}

\begin{figure}[h!]
    \begin{tikzpicture}
    [
      ->,>=stealth',
      shorten >=1pt,
      node distance=3.5cm,
      initial text=$ $
    ]

    \tikzset{ls/.style={font=\small}}

    \node[state,initial] (0) {$0$};
    \node[state,right of=0,xshift=0.8cm] (1) {$1$};
    \node[state,below of=0] (2) {$2$};
    \node[state,below of=1] (3) {$3$};

    \draw
    (0) edge[loop above] node[ls]{$\var{bid} [1] \:/$} (0)
    (1) edge[loop above] node[ls]{$\var{reveal} [1] \:/$} (1)
    (2) edge[loop below] node[ls]{$\var{unbid} [1] \:/$} (2)
    (3) edge[loop below] node[ls]{$\var{withdraw} [1] \:/$} (3)
    (0) edge[above] node[ls]{$\var{close} [\var{biddingOver}] \:/$} (1)
    (0) edge[above] node[ls,rotate=90]{$\var{cancel} [1] \:/$} (2)
    (1) edge[above] node[ls,sloped]{$\var{cancel} [1] \:/$} (2)
    (1) edge[above] node[ls,below,rotate=90,text width=2cm,align=center]{$\var{finish}$ \\ $[\var{revealOver}] \:/$} (3);
  \end{tikzpicture}
  \caption{Blinded auction. FSM generated for $\var{size} = 4$
    states.}
  \label{fig:blinded-auction}
\end{figure}
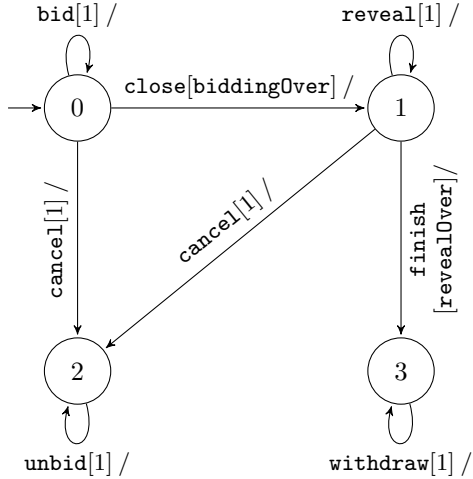

\begin{figure*}[h!]
\begin{lstlisting}[multicols=2]
contract BlindedAuction {
  enum State { ST_0, ST_1, ST_2, ST_3 }
  State private state = States.ST_0;
  struct Bid {
    bytes32 blindedBid;
    uint deposit;
  }
  mapping(address => Bid[]) private bids;
  mapping(address => uint) private pendingReturns;
  address private highestBidder;
  uint private highestBid;
  boolean private closed = false;
  
  function biddingOver() private returns (bool) {
    return now > creationTime + 5 days;
  }  
  function revealOver() private returns (bool) {
    return now >= creationTime + 10 days;
  }

  function cancel() public {
    require(state == States.ST_0
            || state == States.ST_2)
    if (state == ST_0) {
      _cancel_action(); state = ST_3;
    }
    if (state == ST_2) {
      _cancel_action(); state = ST_3;
    }
  }
  
  function      bid() public {...}
  function    close() public {...}
  function withdraw() public {...}
  function   reveal() public {...}
  function   finish() public {...}
  function    unbid() public {...}
  
  function _bid_action() public {
    bids[msg.sender].push(Bid({
      blindedBid: blindedBid,
      deposit: msg.value
    }));
    pendingReturns[msg.sender] += msg.value;
  }

  function    _close_action() public {...}
  function   _reveal_action() public {...}
  function   _finish_action() public {...}
  function _withdraw_action() public {...}
  function   _cancel_action() public {...}
  function    _unbid_action() public {...}
}
\end{lstlisting}
  \caption{Blinded auction. Generated Solidity code.}
  \label{fig:blinded-auction-code}
\end{figure*}

This section contains case studies that evolve from simple example for
illustration purposes only to more realistic examples taken from the
literature. For each case study we provide a formal LTL specification
and a result model that was generated with this specification and test
scenarios.

In this section we extend FSMs with guard conditions.  A \textit{guard
  condition} is a Boolean expression that labels FSM transition.
Guard conditions affect the semantics of FSMs in the following way:
the transition can be executed only if its guard condition is
satisfied.  We use this notation to label transitions:
\mbox{\textit{input event [guard condition] / output action}}.  A test
scenario now is a sequence of triples $\{ (e_i, c_i, A_i) \}$, where
$e_i$ is an input event, $c_i$ is a guard condition and $A_i$ is a
sequence of output actions. If a sequence of output actions is omitted
(as in \mbox{\textit{input event [guard condition] /}} or $(e_i, c_i,
.)$) it is implicitly assumed that it consists of one action with the
same name as the corresponding input event.

In our approach input events of an FSM correspond to methods of a
contract and output actions -- to implementations of those methods.
We specify smart contracts with a set of LTL formulas in terms of
these events and actions.  Formal specification is then combined with
test scenarios and provided as input to \textsc{EFSM-tools} to
generate an FSM that complies with given specification and test
scenarios. Figure~\ref{fig:blinded-auction} shows an example of a
generated FSM. Given the implementation of FSM output actions and the
definitions of used predicates, FSM can be translated to executable
code.  Figure~\ref{fig:blinded-auction-code} shows an example of
generated Solidity code, where lines 2\,--\,12 correspond to
contract's state definition, lines 14\,--\,19 -- to predicate
definition and lines 40\,--\,44 -- to $\mathtt{bid}$ action definition
(other action definitions are omitted for brevity).


\subsection{Crowdfunding}
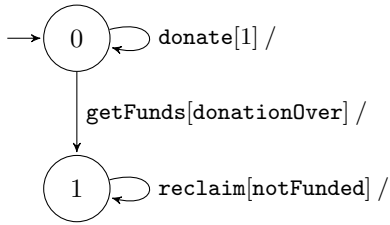
\begin{figure}[h!]
  \centering
    \begin{tikzpicture}
    [
      ->,>=stealth',
      shorten >=1pt,
      node distance=2cm,
      initial text=$ $
    ]

    \tikzset{ls/.style={font=\small}}

    \node[state,initial] (0) {$0$};
    \node[state,below of=0] (1) {$1$};

    \draw (0) edge[right] node[ls]{$\var{getFunds} [\var{donationOver}] \:/$} (1)
    (0) edge[loop right] node[ls]{$\var{donate} [1] \:/$} (0)
    (1) edge[loop right] node[ls]{$\var{reclaim} [\var{notFunded}] \:/$} (1);
  \end{tikzpicture}
  \caption{Crowdfunding. FSM generated for $\var{size} = 2$ states.}
  \label{fig:crowdfunding}
\end{figure}

For illustrative purpose, let us first consider a simplistic example
of a crowdfunding platform.  In such a platform users can donate coins
(denoted with event $\var{donate}$) during a predefined period of time
which ends when variable $\var{donationOver}$ becomes true. When the
donation period is over, the owner of the campaign can request
collected coins (event $\var{getFunds}$). After that donors can claim
their donations back (event $\var{reclaim}$) and possibly get them
back if not enough coins were collected during the campaign
($\var{notFunded} = \true$). Intuitively, one can model the logic
of this contract using an FSM with two states. More formally, the
logic of described contract is formulated as follows:
\begin{enumerate}
\item $\var{getFunds}$ cannot happen more than
  once;\label{prop:crowdfunding.1}
\item $\var{donate}$ cannot happen after $\var{getFunds}$ has
  happened;
\item $\var{reclaim}$ cannot happen before
  $\var{getFunds}$;\label{prop:crowdfunding.3}
\item $\var{getFunds}$ can happen only if $\var{donationOver} =
  \true$;
\item $\var{reclaim}$ can happen only if $\var{notFunded} =
  \true$.
\end{enumerate}
With a straightforward translation we can formalize these properties
in LTL as follows.
\begin{align}
& \textbf{G} (\var{getFunds}\to\textbf{X} \neg \textbf{F} \var{getFunds}) \\
& \textbf{G} (\var{getFunds}\to\neg \textbf{F} \var{donate}) \\
& \var{getFunds}\:\textbf{R}\:\neg \var{reclaim} \\
& \textbf{G} (\var{getFunds}\to\var{donationOver}) \\
& \textbf{G} (\var{reclaim}\to\var{notFunded})
\end{align}
Figure~\ref{fig:crowdfunding} shows an FSM generated with
\textsc{EFSM-tools} from this specification the set of scenarios $S =
\{ s_1, s_2, s_3 \}$, where
\begin{align*}
s_1 = [ & (\var{donate}, \true, .), (\var{donate}, \true, .),\\
        & (\var{donate}, \true, .)];\\
s_2 = [ & (\var{getFunds}, \var{donationOver}, .),\\
  & (\var{reclaim}, \var{notFunded}, .),\\
  & (\var{reclaim}, \var{notFunded}, .)];\\
s_3 = [ & (\var{donate}, \true, .), (\var{getFunds}, \var{donationOver}, .),\\
  & (\var{reclaim}, \var{notFunded}, .), (\var{reclaim}, \var{notFunded}, .)].
\end{align*}
Another advantage of using formal logic is that now we can reason
about about the system.  For instance, from
properties~\ref{prop:crowdfunding.1} and \ref{prop:crowdfunding.3} we
can derive that $\var{getFunds}$ cannot happen after $\var{reclaim}$.

\subsection{Blinded auction}
Now we consider a more realistic example of blinded auction taken
from~\cite{mavridou2017designing}. During a predefined period of time
after contract creation, users can make hidden bids (denoted with
event $\var{bid}$). When this period is over ($\var{biddingOver} =
\true$) the auction can be closed (event $\var{close}$), after
which follows the next period when users are allowed to $\var{reveal}$
their bids. When the second period is over ($\var{revealOver} =
\true$) the auction can be finished (event $\var{finish}$). At
any time before finishing the auction can be canceled
($\var{cancel}$), after which users can claim back their bids (event
$\var{unbid}$).

The logic of this blinded auction can be formulated as follows (we
provide corresponding LTL formulas alongside).
\begin{enumerate}
\item $\var{close}$, $\var{finish}$ and $\var{cancel}$ cannot happen more than once:
  \begin{align*}
    & \textbf{G} (\var{close}\to\textbf{X} \neg \textbf{F} \var{close}), \\
    & \textbf{G} (\var{finish}\to\textbf{X} \neg \textbf{F} \var{finish}), \\
    & \textbf{G} (\var{cancel}\to\textbf{X} \neg \textbf{F} \var{cancel});
  \end{align*}
\item $\var{bid}$ cannot happen after $\var{close}$ has happened:
  \begin{align*}
    & \textbf{G} (\var{close}\to\neg \textbf{F} \var{bid});
  \end{align*}
\item $\var{reveal}$ and $\var{cancel}$ cannot happen after
  $\var{finish}$ has happened:
  \begin{align*}
    & \textbf{G} (\var{finish}\to\neg \textbf{F} (\var{reveal}\vee\var{cancel}))
  \end{align*}
\item $\var{finish}$, $\var{close}$, $\var{bid}$ and $\var{reveal}$
  cannot happen after $\var{cancel}$ has happened:
  \begin{align*}
    & \textbf{G} (\var{cancel}\to\neg \textbf{F} (\var{finish}\vee\var{close}\vee\var{bid}\vee\var{reveal}))
  \end{align*}
\item $\var{finish}$ and $\var{reveal}$ cannot happen before
  $\var{close}$:
  \begin{align*}
    & \var{close}\:\textbf{R}\:(\neg \var{finish}\wedge\neg \var{reveal});
  \end{align*}
\item $\var{unbid}$ cannot happen before $\var{cancel}$:
  \begin{align*}
    & \var{cancel}\:\textbf{R}\:\neg \var{unbid};
  \end{align*}
\item $\var{withdraw}$ cannot happen before $\var{finish}$:
  \begin{align*}
    & \var{finish}\:\textbf{R}\:\neg \var{withdraw};
  \end{align*}
\item $\var{close}$ can happen only if $\var{biddingOver} =
  \true$:
  \begin{align*}
    & \textbf{G} (\var{close}\to\var{biddingOver});
  \end{align*}
\item $\var{finish}$ can happen only if $\var{revealOver} =
  \true$:
  \begin{align*}
    & \textbf{G} (\var{finish}\to\var{revealOver}).
  \end{align*}
\end{enumerate}

This formal LTL specification and a set of test scenarios was used to
generate a FSM depicted in Figure~\ref{fig:blinded-auction}. Test
scenarios for this and the next case study can be found
online~\footnote{\url{https://github.com/d-suvorov/sc-gen}}.  Given
Solidity code associated with output actions labeling FSM transitions,
the full smart contract code can be generated.  The code generated
from synthesized FSM is shown in
Figure~\ref{fig:blinded-auction-code}.

\subsection{License server}
This is an example taken from~\cite{idelberger2016evaluation}. Here we
consider a smart contract that could be used to monitor the execution
of the agreement between two parties, namely Licensor and Licensee. We
assume that these parties perform as client agents connected to some
blockchain network. The contract consists of the following clauses
which were copied from~\cite{idelberger2016evaluation} and annotated
with event names that we are going to use in the rest of this section
to model contract execution.
\begin{enumerate}
\item The Licensor grants the Licensee a license to evaluate the
  Product ($\var{getLicense}$).
\item Licensee must not $\var{publish}$ the results of the evaluation
  ($\var{use}$) of the Product without the approval
  ($\var{getApproval}$) of the Licensor; the approval must be obtained
  before the publication. If the Licensee publishes results of the
  evaluation of the Product without approval from the Licensor, the
  Licensee has 24 hours to $\var{remove}$ the material.
\item The Licensee must not publish comments ($\var{comment}$) on the
  evaluation of the Product, unless the Licensee is permitted to
  publish the results of the evaluation.
\item If the Licensee is commissioned ($\var{getCommission}$) to
  perform an independent evaluation of the Product, then the Licensee
  has the obligation to publish the evaluation results.
\item This license will $\var{terminate}$ automatically if Licensee
  breaches this Agreement.
\end{enumerate}

There is not a timer that can be used to trigger some transition in a
blockchain system, that is why we introduce special events
$\var{noRemove}$ and $\var{noPublish}$. $\var{noRemove}$ happens if
the results of the evaluation of the Product were not removed in due
time. $\var{noPublish}$ happens if the Licensee was commissioned to
perform an independent evaluation and did not published the
results. Thus LTL specification is less abstract and at the first
sight less intuitive than the informal description above. For instance
to state that it is permitted to use and publish results after getting
an approval we introduce the property
$\textbf{G}(\var{getApproval}\to\textbf{G}(\neg\var{noRemove}))$:
``$\var{noRemove}$ cannot happen after $\var{getApproval}$ has
happened''. Given that property, and the fact that $\var{getApproval}$
can only happen after $\var{getLicense}$, we can derive that
$\var{terminate}$ cannot happen after $\var{getApproval}$.  The latter
property can be formalized and verified, which is an advantage of
using formal logic system for specifying a contract. 

We introduce LTL specification in two stages. The following formulas
encode general principles of the system (e.g., ``$\var{remove}$ cannot
happen if nothing has been published'').
\setcounter{equation}{0}
\begin{align}
  & \textbf{G} (\var{getLicense}\to\textbf{X} \neg \textbf{F} \var{getLicense}) \\
  & \textbf{G} (\var{getApproval}\to\textbf{X} \neg \textbf{F} \var{getApproval}) \\
  & \textbf{G} (\var{noRemove}\to\textbf{X} \neg \textbf{F} \var{noRemove}) \\
  & \textbf{G} (\var{noPublish}\to\textbf{X} \neg \textbf{F} \var{noPublish}) \\
  & \var{publish}\:\textbf{R}\:\neg \var{remove} \\
  & \textbf{G} (\var{remove}\to\textbf{X} (\var{publish}\:\textbf{R}\:\neg \var{remove})) \\
  & \var{publish}\:\textbf{R}\:\neg \var{noRemove} \\
  & \textbf{G} (\var{remove}\to\textbf{X} (\var{publish}\:\textbf{R}\:\neg \var{noRemove})) \\
  & \var{getLicense}\:\textbf{R}\:\neg \var{getApproval} \\
  & \var{getLicense}\:\textbf{R}\:\neg \var{getCommission} \\
  & \var{getCommission}\:\textbf{R}\:\neg \var{noPublish} \\
  & \textbf{G} (\var{publish}\to\textbf{X} (\var{getCommission}\:\textbf{R}\:\neg \var{noPublish})) \\
  & \textbf{G} (\var{terminate}\to\textbf{X} \neg \textbf{F} \var{noRemove}) \\ 
  & \textbf{G} (\var{terminate}\to\textbf{X} \neg \textbf{F} \var{noPublish}) \\
  & \textbf{G} (\var{terminate}\to \nonumber\\
  & \phantom{AA}\neg \textbf{F} (\var{getLicense}\vee\nonumber \\
  & \phantom{AA}\phantom{AA}\var{getApproval}\vee\var{getCommission}))
\end{align}
The following formulas encode contractual clauses.
\setcounter{equation}{0}
\begin{align}
& (\var{getLicense}\vee\var{terminate})\:\textbf{R}\:\nonumber\\
& \phantom{AA}((\var{use}\vee\var{publish})\to\var{terminate}) \\
& \textbf{G} (\var{getLicense}\to \nonumber\\
& \phantom{AA}\neg \textbf{F} ((\var{use}\vee\var{publish})\wedge\var{terminate})) \\
& (\var{getApproval}\vee\var{getCommission}\vee\var{terminate})\:\textbf{R} \nonumber\\
& \phantom{AA} (\var{comment}\to\var{terminate}) \\
& \textbf{G} (\var{noRemove}\to\var{terminate}) \\
& \textbf{G} (\var{noPublish}\to\var{terminate}) \\
& \textbf{G} (\var{getApproval}\to\textbf{G} \neg \var{noRemove}) \\
& \textbf{G} (\var{getCommission}\to\textbf{G} \neg \var{noRemove})
\end{align}

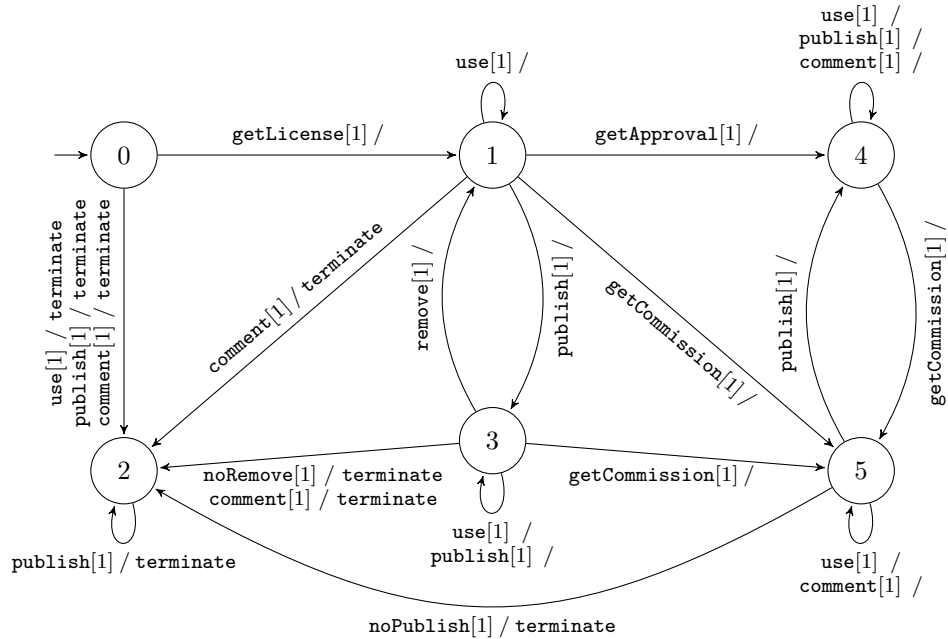
\begin{figure*}[h!]и
  \centering
    \begin{tikzpicture}
    [
      ->,>=stealth',
      shorten >=1pt,
      node distance=3.5cm,
      initial text=$ $
    ]

    \tikzset{ls/.style={font=\footnotesize}}

    \node[state,initial] (0) {$0$};
    \node[state,right of=0,xshift=1.4cm] (1) {$1$};
    \node[state,below of=0,yshift=-0.7cm] (2) {$2$};
    \node[state,right of=2,xshift=1.4cm,yshift=0.4cm] (3) {$3$};
    \node[state,right of=1,xshift=1.4cm] (4) {$4$};
    \node[state,below of=4,yshift=-0.7cm] (5) {$5$};

    \draw
    (1) edge[loop above] node[ls]{$\var{use} [1] \:/$} (1)
    (2) edge[loop below] node[ls]{$\var{publish} [1] \:/\: \var{terminate}$} (2)
    (3) edge[loop below] node[ls,text width=4cm,align=center]{$\var{use} [1] \:/$\\$\var{publish} [1] \:/$} (3)
    (4) edge[loop above] node[ls,text width=4cm,align=center]{$\var{use} [1] \:/$\\$\var{publish} [1] \:/$\\$\var{comment} [1] \:/$} (4)
    (5) edge[loop below] node[ls,text width=4cm,align=center]{$\var{use} [1] \:/$\\$\var{comment} [1] \:/$} (5)
    (0) edge[above] node[ls]{$\var{getLicense} [1] \:/$} (1)
    (0) edge[above] node[ls,above,rotate=90,text width=4cm,align=center]{$\var{use} [1] \:/\: \var{terminate}$\\$\var{publish} [1] \:/\: \var{terminate}$\\$\var{comment} [1] \:/\: \var{terminate}$} (2)
    (1) edge[above,bend left] node[ls,below,rotate=90]{$\var{publish} [1] \:/$} (3)
    (3) edge[above,bend left] node[ls,rotate=90]{$\var{remove} [1] \:/$} (1)
    (3) edge[above] node[below,ls,text width=4cm,align=center,xshift=0.2cm]{$\var{noRemove} [1] \:/\: \var{terminate}$\\$\var{comment} [1] \:/\: \var{terminate}$} (2)
    (1) edge[above] node[ls,sloped]{$\var{getApproval} [1] \:/$} (4)
    (1) edge[above] node[ls,above,sloped]{$\var{comment} [1] \:/\: \var{terminate}$} (2)
    (1) edge[above] node[ls,below,sloped,xshift=0.3cm]{$\var{getCommission} [1] \:/$} (5)
    (4) edge[above,bend left] node[ls,below,rotate=90]{$\var{getCommission} [1] \:/$} (5)
    (5) edge[above,bend left] node[ls,rotate=90]{$\var{publish} [1] \:/$} (4)
    (5) edge[bend left,looseness=1.2] node[ls,below]{$\var{noPublish} [1] \:/\: \var{terminate}$} (2)
    (3) edge[above] node[ls,below,xshift=-0.2cm]{$\var{getCommission} [1] \:/$} (5);
  \end{tikzpicture}
  \caption{License. FSM generated for $\var{size} = 6$ states.}
  \label{fig:license}
\end{figure*}

An FSM generated from this specification and test scenarios is shown
in Figure~\ref{fig:license}. Test scenarios are available online (the
link was provided in the previous section).

The original contract is formulated in terms of deontic modalities,
i.e., in terms of permissions, obligations and related concepts.  A
shortcoming of specifying this smart contract in temporal logic (and
using it to synthesize FSM model) is not tracking these modalities:
given a sequence of actions there is no easy way to figure out
permissions and obligations of contractual parties. On the other hand,
the resulting representation is efficient, which is important in case
of on-chain deployment, and could be used to determine whether or not
the given sequence of actions leads to contract termination.

\section{Conclusion and discussion}
We argue that automated program synthesis could find more applications
for smart contract generation as they often have simpler structure
than general purpose programs and it is an open research question
whether a Turing-complete language is necessary for smart contract
programming. We would like to draw attention of the community to the
problem of automatic synthesis of smart contracts.  We provided
several case studies to show that LTL synthesis can be applied to
generate FSM models for smart contracts of some types. In these models
input events correspond to smart contract methods and output actions
correspond to these methods' implementation. Generated FSM models can
further be used to obtain programs that are correct with respect to
some formal temporal properties.

Our approach can be straightforwardly extended for systems of
interacting smart contracts and used to specify, synthesize and verify
them. Another interesting method to extend the supported class of
verified properties is to incorporate source-level formal verification
techniques. Output actions in synthesized FMSs correspond to smart
contract methods, hence we can use other verification frameworks to
prove source-level properties about these methods and combine them
with temporal properties of FSM models itself.

In practice smart contracts receive, hold and send coins and
transaction execution costs some amount of gas.  It is important to be
able to use these concepts to formulate properties of interest about
smart contracts.  Hence, other possibilities for future work include
using SAT or SMT to encode such concepts and extending specification
language to incorporate these.

A drawback of using LTL synthesis is that LTL is not expressive enough
to formulate properties of kind ``$\phi$ \emph{can happen} infinitely
often (while $\psi$ has not happened)''. For example the formula
$\textbf{EG}\:\phi$ in CTL can be used to state that there is a path
on which $\phi$ always holds. However this problem can be easily
mitigated by specifying test scenarios in which $\var{donate}$ repeats
$N$ times, where $N$ is greater than the number of transitions of FSM
to be generated.

Despite the simple remedy for the above problem we believe that a more
suitable formal system to specify smart contracts is yet to be
identified. It is an interesting research question: how to strike a
balance between simplicity and expressivity of this system to allow
effective synthesis of practical smart contracts.


\ifCLASSOPTIONcompsoc
  \section*{Acknowledgments}
\else
  \section*{Acknowledgment}
\fi

The authors would like to thank Igor Buzhinsky and Daniil Chivilikhin
for their feedback.  This work was supported by the Government of
Russia (Grant 08-08).



\bibliographystyle{IEEEtran}
\bibliography{IEEEabrv,bib,../smart-contracts-verification/bib.bib}
%



\end{document}